\documentclass[12pt,preprint]{aastex}
\usepackage{t1enc}
\usepackage{graphicx}
\usepackage{epstopdf}

\newcommand{\ammonia}{NH$_{3}$}

\newcommand{\mi}{$\mu$m}
\newcommand{\kms}{km~s$^{-1}$}
\newcommand{\chisq}{$\chi^2$}

\newcommand{\msun}{$M_{\odot}$}
\newcommand{\lsun}{$L_{\odot}$}
\newcommand{\accratei}{$\dot M_{\rm i}$}
\newcommand{\accrateacc}{$\dot M_{\rm acc}$}
\newcommand{\mstar}{$M_{*}$}
\newcommand{\rext}{$R_\mathrm{ext}$}
\newcommand{\rew}{$R_\mathrm{ew}$}
\slugcomment{Printed \today}
\shorttitle{Study of the dense core ahead of HH 80N}
\shortauthors{Masqu\'e et al.}

\begin{document}

\title{A multi-wavelength study of the star-forming core ahead of HH 80N}

\author{Josep M. Masqu\'e\altaffilmark{1}, Mayra Osorio\altaffilmark{2}, Josep M.
Girart\altaffilmark{3}, Guillem Anglada\altaffilmark{2}, Guido Garay\altaffilmark{4}, Robert
Estalella\altaffilmark{1}, Nuria Calvet\altaffilmark{5} \and Maria T. Beltr\'an\altaffilmark{6}}

\altaffiltext{1}{Departament d'Astronomia i Meteorologia, Universitat de Barcelona, 
Mart\'i i Franqu\`es 1, 08028 Barcelona, Catalunya, Spain}

\altaffiltext{2}{Instituto de Astrof\'isica de Andaluc\'ia, CSIC, Glorieta de la Astronom\'ia S/N,
E-18008 Granada, Spain}

\altaffiltext{3}{Institut de Ci\`encies de l'Espai, (CSIC-IEEC), 
Campus UAB, Facultat de Ci\`encies, Torre C5 - parell 2, 
08193 Bellaterra, Catalunya, Spain}

\altaffiltext{4}{Departamento de Astronom\'ia, Universidad de Chile, Camino el
Observatorio 1515, Las Condes, Santiago, Chile}

\altaffiltext{5}{Department of Astronomy, University of Michigan, 830 Dennison Building, 500 Church Street, Ann Arbor, MI 48109, USA}

\altaffiltext{6}{INAF - Osservatorio Astrofisico di Arcetri, Largo E. Fermi 5, 50125 Firenze, Italy}

\begin{abstract}

We present observations of continuum emission in the MIR to mm wavelength 
range, complemented with ammonia observations, of the dense core ahead of 
the radio Herbig Haro object HH 80N, found in the GGD 27 region. The 
continuum emission in all the observed bands peaks at the same position, 
consistent with the presence of an embedded object, HH 80N-IRS1, within 
the core. The distribution of the VLA ammonia emission is well correlated 
with that of the dust, suggesting that photochemical effects caused by the 
nearby Herbig Haro object do not play an important role in shaping this 
particular molecular emission. In order to unveil the nature of HH 
80N-IRS1 we analyzed the continuum data of this source, using 
self-consistent models of protostellar collapse. We find that a young 
protostar surrounded by a slowly rotating collapsing envelope of radius 
$\sim$0.08~pc and 20~\msun\ plus a circumstellar disk of radius 
$\sim$300 AU and 0.6~\msun\ provide a good fit to the observed spectral 
energy distribution and to the maps at 350~\micron, 1.2~mm and 3.5~mm of 
HH 80N-IRS1. Besides, the APEX and PdBI continuum maps at 350~\micron\ and 
3.5~mm, respectively, reveal additional clumps in the continuum emission. 
Given the modeling results and the observed morphology of the emission, we 
propose a scenario consisting of a central embedded Class 0 object, HH 
80N-IRS1, with the rest of material of the HH 80N core possibly undergoing 
fragmentation that may lead to the formation of several protostars.

\end{abstract}
%\today

\section{Introduction}

The region of GGD 27 (Gyulbudaghian, Glushkov, \& Denisyuk 1978) in 
Sagittarius at a distance of 1.7~kpc (Rodr\'iguez et al. 1980), is an 
active star forming region with many properties still unveiled. The most 
well-known observational signature of this region, the HH 80/81/80N jet, 
is one of the largest collimated jet systems known so far, expanding over 
a total length of $\sim$~5 pc (Mart\'i et al. 1993). Synchrotron 
radiation, indicating the presence of relativistic electrons, has been 
found in this jet (Carrasco-Gonz\'alez et al. 2010).  The jet is powered 
by a young, high-luminosity protostellar object, IRAS 18162$-$2048. 
Recently, Fern\'andez-L\'opez et al. (2011) detected compact millimeter 
emission towards this source, that was interpreted by these authors as 
arising from a massive ($\sim4$~\msun), and compact ($r \la 300$~AU) disk. 
Copious UV radiation, generated in the Herbig Haro (HH) shocks and the jet, is able to 
induce the formation of a Photodissociation Region (PDR) along the bipolar 
flow (Molinari et al.\ 2001).

Ahead of the radio source HH 80N, the obscured northern head of the jet, 
there is a dense core of $\sim$0.3 pc in size (hereafter HH 80N core) 
first detected in ammonia (see Fig.~1 of Girart et al. 1994)  and 
afterward in other molecular species (Girart et al. 1998, 2001; Masqu\'e 
et al. 2009). A comparison between the CS, NH$_3$ and HCO$^+$ emission led 
Girart et al (1998) to suggest an unusual chemistry for this core. From 
the ammonia emission of the core a mass of roughly 20~\msun\ was estimated 
by Girart et al. (1994).

Claims for evidence of association of molecular clumps with HH objects  have been
growing over the last decades (Rudolph \& Welch\ 1988; Davis,  Dent \& Bell Burnell
1990; Torrelles et al.\ 1992, 1993; Girart et al.\  1994, 1998; 2002, 2005; Viti,
Girart \& Hatchell 2006; Whyatt et al.\  2010). These clumps do not show evidence
for shock heating as indicated by  their typically observed narrow linewidths
(<~1~\kms), low temperatures  (<~20~K) and radial velocities close to the ambient
cloud velocity (e.g.  Whyatt et al.\ 2010). All these properties rule out the
possibility of a dynamical perturbation. A possible explanation for the
association of molecular clumps with HH objects, proposed by Wolfire  \& K\"{o}nigl
(1992), is that the molecular clump is exposed to a strong UV  radiation generated
in the HH objects. This radiation is capable of  evaporating icy mantles on dust
grains, triggering a non-equilibrium  chemistry that leads to an increase in
abundance of specific molecular  species. This scenario was extensively modeled by
Taylor \& Williams (1996) and Viti \& Williams (1999).

In the frame of the photo-illuminated scenario, Girart et al. (1998)  suggest
that the observed chemistry of the HH 80N core is compatible with  the
radiative shock-induced chemistry models, being HH 80N the most  plausible
source of UV irradiation. A more recent study, using several  molecular
species, appears to be in agreement with this chemical scenario,  even though
the photochemical effects are observed tentatively only in the  side of the
core facing HH 80N (Masqu\'e et al. 2009). However, the mass and size of the HH
80N core make this clump quite  different from the other molecular clumps found
ahead of HH objects. The latter are smaller and less massive (size~$\le$~0.1
pc; $M \simeq  1~M_{\odot}$) than the HH 80N core and do not show any signpost
of star  formation. Indeed, they are believed to be transient structures or
small  density fluctuations within the molecular clouds (Viti et al. 2003;
Morata  et al. 2003, 2005).

Interestingly, Girart et al. (2001) found evidence of a bipolar CO outflow 
centered near the peak of the HH 80N molecular core, suggesting the  presence of
an embedded protostar. In addition, Girart et al. (2001) and  Masqu\'e et al.
(2009) interpreted the distribution and kinematics of CS  and other molecular
species observed in the outer parts of the HH 80N core  as suggestive of an
infalling ring-like molecular structure (probably  caused by strong molecular
depletion in the inner parts) with a radius of  $\sim$0.24 pc. The infall
velocities inferred ($\sim 0.6$ \kms), significantly larger  than those expected
in the standard protostellar collapse (e.g. ambipolar  diffusion models predict
gas inward motions of a fraction of the  isothermal sound speed, $\sim$0.2~\kms,
at scales similar those of the HH  80N core; Basu \& Moscouvias 1994), led these
authors to suggest a  peculiar dynamical evolution of this core. Given the
particular  environment of the HH 80N core, we question whether the HH 80/81/80N
outflow has triggered or at least sped up in some way  the collapse of the core.

In order to unveil the nature of the HH 80N core, we carried out an 
extensive set of continuum and ammonia observations over the last years 
using several instruments, among them VLT and APEX. Using radiative 
transfer models of collapsing protostellar envelopes, we compare the 
synthetic spectral energy distribution (SED) and spatial intensity 
profiles of the dust emission with the data. From this comparison we 
derive the mass and luminosity of the collapsing envelope, mass infall 
rate, and mass of the embedded object under the hypothesis that a young 
stellar object (YSO) is forming inside of the HH 80N core. We explore 
several density profiles for the initial configuration of the collapsing 
envelope, and discuss whether the inclusion of an accretion disk is needed 
to simultaneously explain the observed emission at different wavelengths. 
Our results point to a new picture for the HH 80N core, which consists in 
a 'classic' Class 0 object (hereafter, HH 80N-IRS1) surrounded by a 
reservoir of material where other clumps potentially forming stars are 
found.

The paper is organized as follows.  In Section~2, we describe the 
observations. In Section~3 we present the resulting maps. In Section~4 we 
model the observed dust emission using different approaches for the 
protostellar collapse. The discussion of the results of the modeling is 
done in Section 5. Finally, our conclusions are summarized in Section~6.

\section{Observations}

\subsection{VLA}

We observed the (J,K) = (1,1) and (2,2) ammonia inversion transitions (at 23.6944955~GHz and
23.7226336~GHz, respectively) using the Very Large Array (VLA) of the NRAO\footnote{ The National
Radio Astronomy Observatory is a facility of the National Science Foundation operated under
cooperative agreement by Associated Universities, Inc.} in the D configuration on January 31, 2007.
The 4IF spectral correlator mode was used, which allows to observe both transitions in two
polarizations simultaneously.  The correlator was set to observe a bandwidth of 1.56 MHz with 63
spectral channels of 24.4 kHz (which gives 0.115~\kms\ of velocity resolution at 1.3 cm) plus a
continuum channel that corresponds to 75\% of the bandwidth. The phase center of the observations
was set to $\alpha(J2000)=18^\mathrm{h}19^\mathrm{m}18\fs81$ and
$\delta(J2000)=-20\arcdeg40'55\farcs0$. The flux calibrator was 3C286 with a flux of 2.59~Jy, the
phase calibrator was 1832-105 with a bootstrapped flux of $0.88 \pm 0.01$ Jy and the bandpass
calibrator was 0319+415. Maps of the ammonia (1,1) transition were obtained with natural weighing
and using a Gaussian taper of 35~k$\lambda$, which gives a beam size of $6.3\arcsec \times
4.4\arcsec$ (P.A. = 20.0$\arcdeg$) and an {\em rms} noise level of $5 \times 10^{-3}$~Jy~beam$^{-1}$
{\em per} channel. Here we only present the map of the velocity-integrated flux density of the (1,1)
transition, which presents extended emission with a good signal to noise ratio (see \S~3.2),
for a comparison with the continuum data. A more complete analysis including the ammonia (2,2)
transition, that is sensitive to warmer gas, will be presented in a future paper.

\subsection{IRAM 30m}

The 1.2~mm continuum observations were carried out with the 117-channel bolometer array
(MAMBO~2) installed at the 30m IRAM telescope (Pico Veleta, Spain) on January 27
and 28 and March 3, 2008. The data were taken under low sky-noise conditions with the
zenith opacity at 250~GHz ranging from 0.14 to 0.28. The on-the-fly technique was used with
a scanning speed of $4\arcsec$~s$^{-1}$. Chopping was performed with throws of
$58\arcsec$ and $70\arcsec$. The angular resolution of the map is $10\farcs5$
and the achieved rms noise level at this angular resolution is
1.5~mJy~beam$^{-1}$. Data reduction was carried out with the MOPSIC software.

\subsection{Plateau de Bure}

The Plateau de Bure Interferometer (PdBI) observations were carried out on April 3 and 10, 2010 in the C
configuration. Two tracks were performed under good weather conditions.  We made a 6-point mosaic with
the phase tracking center of the observations set at $\alpha(J2000)=18^\mathrm{h}19^\mathrm{m}17\fs81$ and
$\delta(J2000) = -20 \degr 40\arcmin 47\farcs7$, coincident with the Spitzer 8~\micron\ peak position (see
\S~3.1). The receivers were tuned to the rest frequency of the N$_2$D~(1$_{1,1}$--1$_{0,1}$) line. The
correlator was configured in  four widex units (two units for each polarization) that gives a total
continuum bandwidth of 8~GHz, plus two narrow windows of 20~kHz covering the N$_2$D~(1$_{1,1}$--1$_{0,1}$)
(85.926263~GHz) and HN$^{13}$C~(1--0) (87.090859~GHz) transitions, with a velocity resolution of $\sim
0.15$~\kms.   The phase calibrator was QSO~1911-201, which has a flux density of 1.09~Jy at the observed
frequency,  and the flux and bandpass calibrator was MWC349. The data were calibrated using CLIC following
the baseline-based mode, since antenna-based solutions were not optimal for the shortest baselines. The
continuum map was obtained using MAPPING with natural weighting, that gives a synthesized beam (HPBW)
of $7.0\arcsec \times 2.9\arcsec$ (P.A. = $11\arcdeg$) at 86~GHz. The spectral line data will be presented
in a future paper.

\subsection{VLT}

In the nights of May 2 and June 12, 2009, Q-band (20~\micron) imaging of the HH 80N
region was carried out in service mode using the VISIR instrument installed at the
Cassegrain focus of the UT3 telescope (Melipal) of the Very Large Telescope (VLT).
The observations covered a field of view of $32.5\arcsec \times 32.5\arcsec$ at a
pixel scale of $0.127\arcsec$/pixel. 

The sky conditions were good: for the night of May 2 the optical seeing
ranged from $0\farcs8$ to $0\farcs9$ and the air-mass was $\sim$~1.05 in average;
for the night of June 12 the average optical seeing varied from $1\farcs0$ to
$1\farcs5$ and the air-mass was always less than 1.05. Under these conditions,
the extrapolated seeing at Q band according to the Roddier formula ($\propto
\lambda^{0.2}$) is $< 0\farcs5$. Taking into account that the VLT diffraction
limit at the same band is $0\farcs5$, this implies that the angular resolution
of our Q band data is basically dominated by diffraction.      

During the first observing night our target was observed for one hour through the
Q1, Q2, and Q3 filters (covering  $17.65\pm0.83$~\micron, $18.72\pm0.88$~\micron,
and $19.50\pm0.40$~\mi, respectively). However, due to the low luminosity of the
target, it was not detected in any of these filters. In the second night, our
target was observed solely in the Q2 filter for one hour and was detected with a
signal-to-noise ratio of 17.  In order to remove the atmospheric and telescope
background the standard chopping and nodding technique in perpendicular directions
was carried out with chop-throws of $14\arcsec$.  The standard star for
photometric calibration, HD178345 (3.52, 3.15 and 2.88~Jy through Q1, Q2 and Q3
filters, respectively), was observed immediately after our target. A preliminary
reduction of the data was carried out using the standard ESO reduction software
including the graphical user interface to the pipeline, GASGANO. The final image
resulting of shifting and combining the chopping and nodding cycles was obtained
with the IRAF package. 

The photometric calibration was performed using the 'visir\_img\_phot' recipe with the
combined image of the cataloged standard star as an input. We obtained conversion factors
of 11923.3~(Q1), 11235.0 (Q2, first night), 13379.4~(Q2, second night) and 2446.2~(Q3)
between the number of detector counts per second and the source flux in Jy. Q2-band flux
density of the target was obtained with standard aperture photometry using the PHOT task
of IRAF with a circular aperture of radius $1\farcs9$ although we inspected values between
$1\farcs5$ and $2\farcs5$. For the background subtraction, the sky contribution was fitted
to an annulus situated between radii of $1\farcs9$ and $2\farcs8$ from the center of the
aperture. Applying the conversion factor derived above, we obtain a flux density value of
0.175~mJy at 18.7 \micron. The flux uncertainty was derived exploring the variation of the flux density value when
measured using different apertures.

\subsection{APEX}

The sub-mm data were obtained with the SABOCA camera, a 39-pixel bolometer array
located on the Atacama Pathfinder EXperiment (APEX) telescope in the Chilean Andes.
Each SABOCA pixel consists in a composite bolometer with superconducting thermistor on
silicon-nitride membranes. The pixels are arranged in a hexagonal layout consisting of a
central channel and 3 concentric hexagons. The array is installed at the Cassegrain focus,
where it has an effective field of view of $90\arcsec$. SABOCA operates at 850 GHz
(350~\micron) which gives a beamsize (FWHM) of $7\farcs7$. The observations were carried out on October 7, 2009 in raster spiral mode with 4 scans, each
providing a fully sampled area of $\sim120\arcsec \times 120\arcsec$. Two  skydips, taken
in-between and after on-source observations, yielded values of $\sim$~1.1 for the atmospheric
zenith opacity. Pointing was checked on secondary calibrators G10.62, G5.89 and HD-163296.

All the scans were reduced using the miniCRUSH software, a reduced version for the APEX bolometers of the SHARC-2 data reduction package CRUSH. We used the default
reduction procedure and the data were smoothed using a $4\farcs0$ Gaussian that gives an
angular resolution of  $8\farcs5$ and an {\em rms} noise of 0.09~Jy~beam$^{-1}$ for the final map.  

\subsection{Infrared Archive Data from IRAS, Spitzer and Akari}

The IRAS Point Source Catalog (PSC) reports a weak source, IRAS 18163-2042, 
whose position uncertainty includes the HH 80N core. According to the catalog,
this source was detected only at 60~\mi\ with a flux density of $\sim$~5~Jy and
remained undetected in the rest of the bands ($\le$~0.5~Jy at 12 and 25~\mi, and
$\le$~250~Jy at 100~\mi). However, these values are not reliable because of the
presence of strong side lobes generated by the nearby luminous source IRAS
18162-2048. These side lobes create a complex background around IRAS 18163-2042,
which may cause an underestimate of the flux values reported in the IRAS PSC.
This motivated us to SCANPI reprocess the four IRAS bands covering the HH 80N
region. SCANPI, a utility provided by the Infrared Processing and Analysis 
Center (IPAC), performs 1-dimensional scan averaging of the IRAS raw survey
data.  As it combines all the scans passed over a specific position, its
outcome  has a higher sensitivity, ideal to obtain fluxes of confused or faint
sources. In particular, among all the possible input processing parameters of
SCANPI, we gave special attention to the `local background fitting range'. By
default, an interval of radius $60'$ from the scan center is used to fit the
background. From this interval, a central range depending on the band is
excluded in the fitting in order to prevent contamination from the target: $2'$
(12~\mi\ and 25~\mi), $4'$ (60~\mi) and $6'$ (100~\mi).  Using this default
SCANPI yields values similar to the fluxes of the IRAS PSC for all the 1-D scans
that pass thorough the HH 80N region. However, IRAS 18162$-$2048 is located $6'$
south-east from IRAS 18163$-$2042 and, hence, the default range for the
background subtraction includes clearly the emission of this luminous source.
Adopting $12'$ for the `source exclusion range' in the background fitting for
all the bands, we obtained significantly higher flux values than those derived
above. Other values for the `source exclusion range' yielded similar results
provided that IRAS 18162$-$2048 is excluded from the background fitting range.

To determine better the flux at IRAS bands we checked the Summary Tables in the Result
Details page within the SCANPI website. The Summary Tables present single scan data. From
this data, we plotted the flux after background subtraction vs. sky offset for each single
scan and inspected manually each plot in order to select the scan that showed the best
detection. From the selected scan we adopted as the flux value the maximum value of the
plot found within the offset interval passing through the target position. We performed
this process for all the IRAS bands and obtained new flux values.

We also retrieved observational data of the HH 80N region from the Spitzer archive at 4.5 and 8~\micron\
IRAC bands that are part of a large sample of high-mass star forming regions  (PID: 3528). The data at
these wavelengths complement very well our VLT observations.

Following the same procedure as for the Q band data, we estimated the flux density
at 4.5 and 8~\micron\ with the aperture photometry technique using PHOT of IRAF.
We found that significantly larger apertures must be used  for the Spitzer data 
compared to the apertures used for the VLT data: $11\arcsec$ and $12\arcsec$ for
and 4.5 and 8~\micron\ images, obtaining flux values of 0.026 and 0.041~Jy for 4.5
and 8~\micron, respectively.  As for the VLT data, the flux uncertainty was
derived by monitoring the variation of the flux value when measured using
different apertures. 

Finally, we made use of the recently available data from the Infrared Astronomical Satellite
Akari. Akari is equipped with two instruments, IRC, covering several bands at near  and
mid-infrared wavelengths, and FIS, covering several bands at far-infrared wavelengths. From a
total of six observable bands, in this paper we present the data of  18, 140 and 160~\micron\
bands. The rest of bands have bad quality data or null detection for our source.

\section{Results}

\subsection{Continuum emission}

\begin{figure}[fh]
\epsscale{0.5}
\plotone{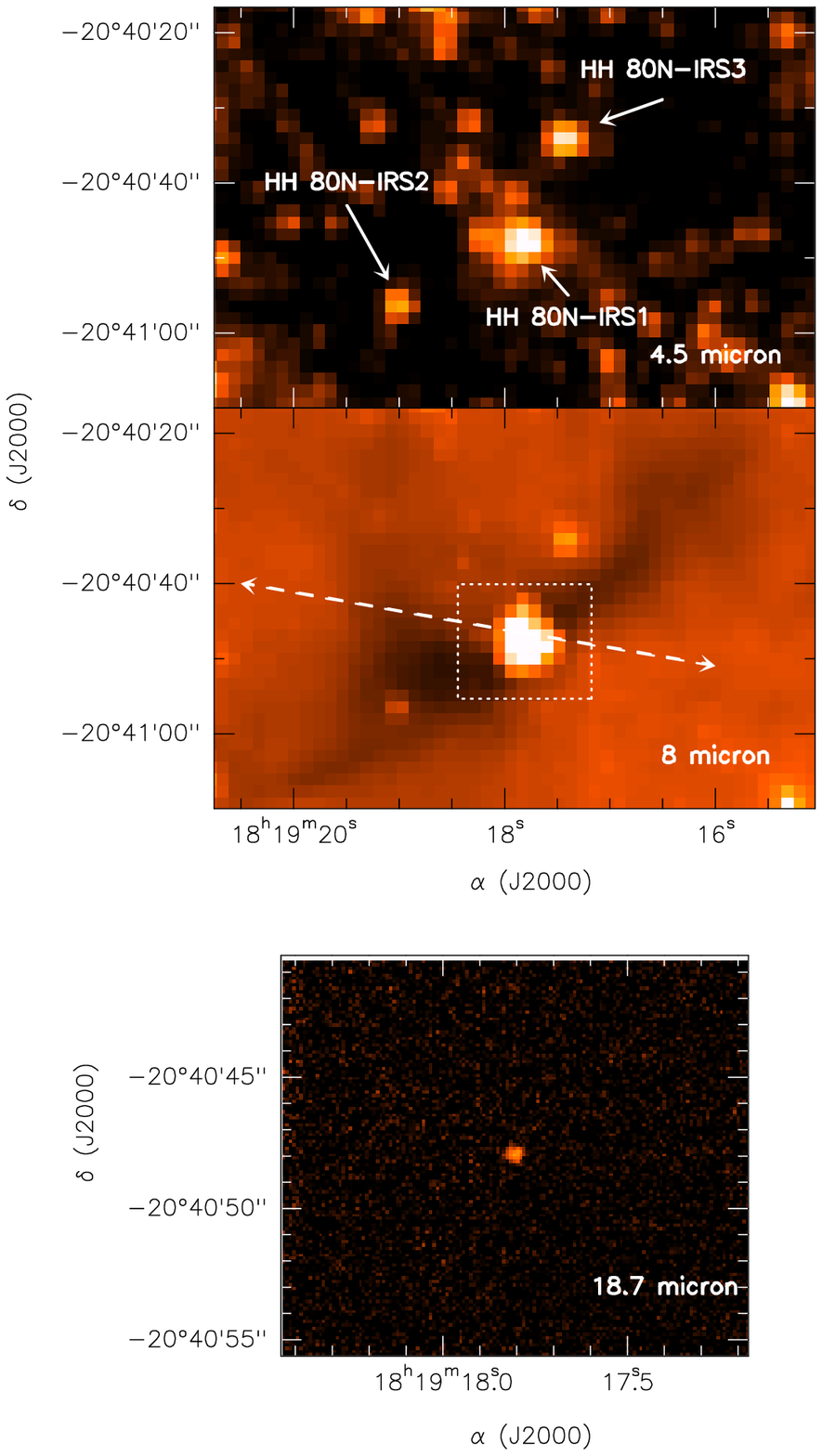}
\vspace{0.0 cm}
\caption{
Images of the HH 80N region at  4.5~\micron\  (top panel) and 8~\micron\ (middle panel), retrieved from the Spitzer
archive, and 18.7~\micron\  (bottom panel) obtained with the VLT. In the middle panel, 
the dashed square shows the limits of the VISIR (VLT) field of view and the
dashed arrow gives approximately the orientation and extend of the outflow 
detected by Girart et al.\ 2001.\label{cont_MIR}}
\end{figure}

Figure~\ref{cont_MIR} presents the mid-IR images of the HH 80N region. The Spitzer images at 4.5
and 8~$\mu$m show a bright compact source dominant in the two wavelengths (HH 80N-IRS1). Two other
compact sources, located $\sim 20''$ South-East (HH 80N-IRS2) and $\sim 15''$ North-West (HH
80N-IRS3) of HH 80N-IRS1, may belong to the region. The positions of  these compact sources are
given in Table~\ref{spitzer}. HH 80N-IRS1 is also detected in the VLT image at 18.7~\micron, but
without any extended structure, probably due to the short exposure time of our observations. 
Taking into account that the source is not resolved, and given the angular resolution of VLT at
Q-band  ($\sim 0\farcs5$), the warm part of the envelope remains within $\sim$~425~AU (for 1.7~kpc
of distance) from the core center. 

HH 80N-IRS1 is located at the center of the HH 80N core that is seen in absorption against the
emission of the Galactic background in the 8~\micron\ image. This observational picture
resembles that of InfraRed Dark Clouds (IRDC), being the  size of the silhouette of the HH 80N
core ($\sim$~0.4~pc) comparable to the lower limit of the range derived for a sample of IRDCs
(0.4-15~pc, Carey et al.\ 1998). However, while IRDCs harbor the earliest evolutionary stages of
high mass star formation (Rathborne et al.\ 2010), the mass estimated for the HH 80N core (see \S~1 and below) seems too low to identify this core as a potential site for the formation of 
high mass protostars.

\begin{figure}[fh]
\epsscale{0.5}
\plotone{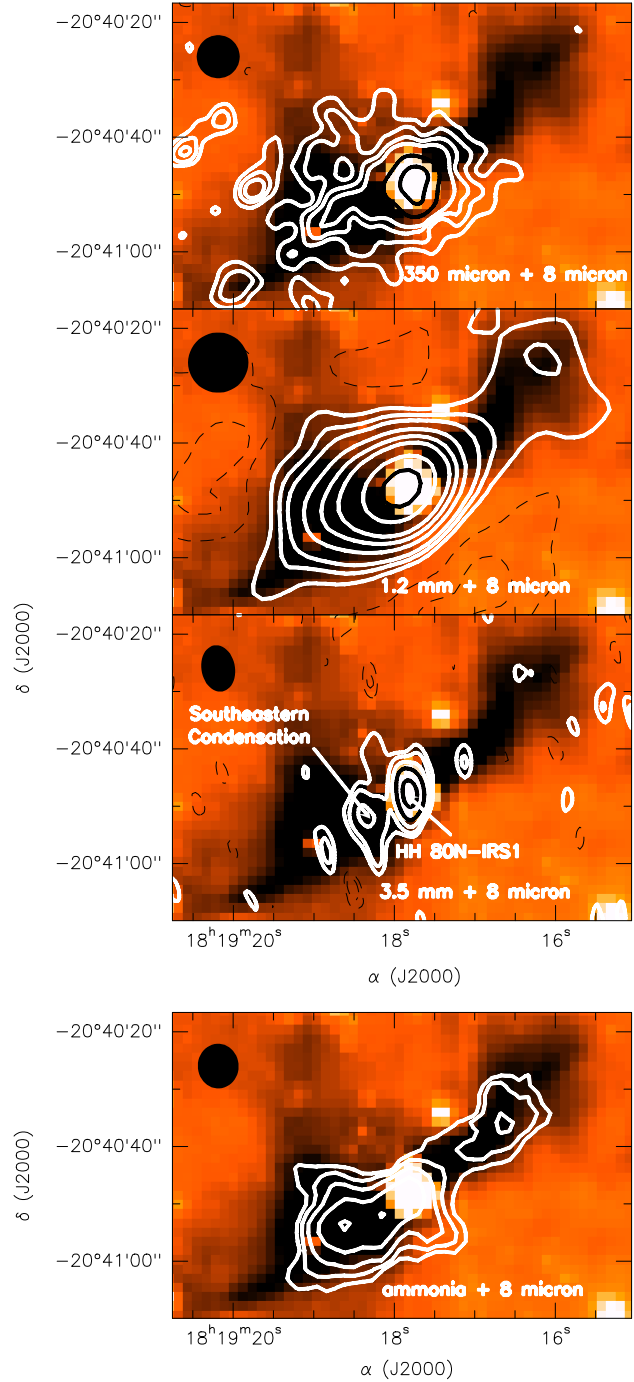}
\vspace{0.0 cm}
\caption{
Maps of the HH 80N region taken at 350~\micron\ (top panel), 1.2~mm (second panel),  3.5~mm (third panel) and the velocity
integrated (zero-order moment) of the main line of \ammonia (1,1) emission (bottom panel), superimposed on the Spitzer 8 $\mu$m image. 
Contour levels are 3, 4, 6, 9, 15, and 21 times 90~mJy~beam$^{-1}$ (350~\micron); $-$6, $-$3, 3, 6, 9, 15, 21, 27, 39, and 50
times 1.5~mJy~beam$^{-1}$ (1.2~mm); $-$3, $-$2, 2, 3, 5, 7 and 10 times 0.11~mJy~beam$^{-1}$ (3.5~mm); 3, 6, 10, 16 and 24
times 0.24~Jy~beam$^{-1}$~\kms (\ammonia). The beams are shown in the upper left corner of the panels. The color scale of the
infrared image has been modified with respect to Fig.~\ref{cont_MIR} in order to highlight the absorption feature (see text).
\label{cont_ammonia}}
\end{figure}

As seen in Figure~\ref{cont_ammonia}, the 350~\micron\ emission is  elongated with an angular size
of $20\arcsec \times 15\arcsec$ (FWHM) and  P.A.~$\sim$~120$\arcdeg$, and peaks at the position of
HH 80N-IRS1. Apart  from HH 80N-IRS1, the 350~\micron\ emission traces additional material  towards
the South-East. The 1.2~mm emission peaks at the same position and  traces fairly well the
silhouette of the absorption feature of the  8~\micron\ image including the north-western tail
expanding up to  $35\arcsec$ from the central peak. The detection of this tail in emission  at
1.2~mm and in absorption at 8~\micron\ excludes the possibility of  being an artifact of the 1.2~mm
map.  In the PdBI 3.5~mm map the dust  emission splits into two main sources, one clearly associated
with HH  80N-IRS1, plus another southeastern component (hereafter Southeastern  Condensation).
Table~\ref{sources} gives the results of Gaussian fits of  the 3.5~mm emission for HH 80N-IRS1 and
the Southeastern Condensation.   In addition, the 3.5 mm map shows two marginally detected sources
located  $\sim 10\arcsec$ northwest and $\sim 20\arcsec$ southeast of HH 80N-IRS1.

Table~\ref{flux} gives a summary of the flux density measurements towards  HH 80N-IRS1.
Since at mm and submm wavelengths it is difficult to  discriminate which fraction of the
emission of the HH 80N core corresponds  to HH 80N-IRS1, in the table we report a range
of possible values for the  flux densities at these wavelengths. The upper limit of the
range at  1.2~mm and 350~\micron\ is an estimate of the HH 80N-IRS1 flux density 
avoiding the contamination from the Southeastern Condensation. To do this, we  integrated
the flux density of the western half of the HH 80N core and  multiplied the resulting
value by 2. The lower limit is the intensity peak  that would coincide with the flux
density of HH 80N-IRS1 if it were an  unresolved source (i.e. the lowest possible
contribution). For the 3.5~mm  measurement, we take in account the missing short spacings
of the PdBI. We  made a crude analysis simulating the filtering effects of the $u$-$v$ 
coverage of our PdBI observations. Using the UVMODEL task of MIRIAD  package, we tested
these filtering effects on several synthetic maps of  artificially generated ellipses
that mimic the HH 80N core appearance. We  find that a maximum of a 50\%\ of the total
flux is missed. Thus, in the  range of flux densities given for the 3.5~mm emission, the
lower limit  corresponds to the flux density measured in the map and the upper limit 
corresponds to this value corrected by a factor of 2. The data obtained  with low angular
resolution (i.e. all the IRAS data and the Akari 140 and  160~\micron\ bands) are likely
contaminated by background sources. Therefore, we considered these fluxes as upper
limits.  Finally, note that the complex background of the 8~\micron\ image (see 
Fig.~\ref{cont_MIR}) makes the estimate of the flux at this wavelength  somewhat
uncertain.

\subsection{The nature of \ammonia\ emission}

Figure~\ref{cont_ammonia} (bottom panel) presents the \ammonia\ (1,1) emission
superimposed to the Spitzer 8~\micron\ image. The ammonia is very well
correlated with the 8~\micron\ absorption feature. It is also well  correlated
with the 1.2~mm emission. This is not the case for other molecular tracers such
as CS or SO presented in previous works (Girart et al. 2001; Masqu\'e et al.
2009), whose emission is significantly more extended ($\sim60\arcsec \times
25\arcsec$). These studies also show that, in general, the molecular tracers do
not peak all at the same position, probably because these molecules are depleted
in the densest and inner part of the HH 80N core, that is well traced by the
dust continuum emission.

The clear correlation between the \ammonia\ emission and dust emission indicates
that \ammonia\ traces fairy  well the material of the HH 80N core. Girart et al.
(2001) detected star forming signatures in the core such as a bipolar outflow traced
by CO. In addition, the dust continuum emission shows a compact source, HH 80N-IRS1,
in  all the observed wavelengths, suggesting the presence of an embedded YSO. All
these results suggest that the HH 80N core is currently undergoing active star
formation, and that the observed distribution of \ammonia\ emission  arises as a
consequence of the high gas densities likely reached in the core, similarly to other
star forming cores, and not as a consequence of photochemical effects. Ammonia
emission arising as a consequence of a dynamical perturbation is excluded by the narrow
\ammonia\ (1,1) linewidth ($\sim$~1~\kms) observed. Nevertheless, we note that the
strongest \ammonia\ emission is found in the Southeastern part of the core, close to
HH 80N, coinciding with the emission detected  in the lower sensitivity ammonia
observations of Girart et al.\ (1994). This could be due to a local increment of
abundance in this part of the core as found in some species (Masqu\'e et al.\ 
2009); indeed, \ammonia\ is one of the species predicted to be enhanced by the HH
radiation (Viti et al.\ 2003). Understanding these local departures of the ammonia
emission from the global distribution of gas and dust in the HH 80N core is an issue
that will require further investigation.

\section{Modeling}

In the following we analyze this region assuming that an YSO is forming  inside the HH 80N core. 
To do that, we calculate the dust emission arising from an envelope of  dust and gas that
is collapsing onto a central star. We consider three possible density  profiles for the
envelope. We first investigate the collapse of a Singular Logatropic Sphere (SLS, 
McLaughlin \& Pudritz 1996; Osorio et al. 1999, 2009). The SLS has a logarithmic
relationship  between pressure and density, introduced by Lizano \& Shu (1989) to
empirically take into account the  observed turbulent motions in molecular clouds. In the
SLS collapse solution an expansion  wave moves outward into the static core and sets the
gas into motion toward the central star.  Outside the radius of the expansion wave the SLS
envelope is static, with a dependence of the  density with radius as $\rho \propto
r^{-1}$. Inside the radius of the expansion wave the gas  falls onto the central star with
a nearly free-fall behavior ($v\propto r^{-1/2}$, $\rho  \propto r^{-3/2}$) at small
radii.

As a second approach, we adopt the collapse solution of the Singular  Isothermal Sphere
(SIS, Shu 1977). In this case the collapse occurs in a similar fashion  than in the
logatropic case but the radial dependence of the density in the static region goes as 
$\rho \propto r^{-2}$. Nevertheless, there are important differences in the evolution of
both  types of collapse as both the speed of the expansion wave and the mass infall rate
are  constant in the SIS collapse  while they  increase with time in the SLS collapse.

As a third approach we use the solution for the collapse of a slowly  rotating core
described in Terebey, Shu and Cassen (1984), hereafter the TSC collapse (see also  Cassen
\& Mossman 1981; Kenyon et al. 1993). In this model, the initial equilibrium state 
corresponds to the uniformly rotating analogue of the SIS. To first order, the collapse
proceeds  similarly to the isothermal case beginning at the center of the core and
propagating outward at the sound speed as an expansion wave. Material outside the radius
of the expansion wave  remains in hydrostatic equilibrium (with $\rho \propto r^{-2}$)
while  inside this radius the infall velocity and density approach those of  free fall.
However, the angular momentum of the infalling gas becomes  important in the vicinity of
the centrifugal radius, where motions become significantly non radial and  material then
falls onto a circumstellar disk rather than radially onto the central  object. The
centrifugal radius is given by $R_c = r_0\Omega_0/(G M_*)$,  where $\Omega_0$ is the
angular velocity at a distant reference radius  $r_0$.

The HH 80N core has a moderate bolometric luminosity, but relatively 
strong mm and submm emission. By integrating the area below the observed 
SED, constructed with the flux densities of Table~\ref{flux}, we can 
derive a possible range of luminosities for HH 80N-IRS1. Considering the 
lower limits of the mm and submm points and excluding the rest of 
continuum data in this calculation, we obtain 10~\lsun\ as the luminosity 
lower limit. Similarly, taking the upper limits of the mm and submm range 
and including the 60~\micron\ IRAS point, which is the most restrictive 
upper limit in the IR part of the SED, we obtain an upper limit of 
110~\lsun\ for the luminosity. We are assuming that the total luminosity, 
which is responsible for internal heating of the core, is the sum of the 
stellar luminosity and the infall luminosity caused by the infalling 
gas onto the protostar.  The stellar luminosity can be related to the mass 
of the central star using the Schaller et al. (1992) evolutionary tracks. 
The upper limit of the luminosity range deduced above (110~\lsun) 
restricts the mass of the central embedded object to $\le$~3~\msun, 
according to the tables of Schaller et al. (1992); this mass upper limit 
corresponds to the hypothetical case that all the luminosity of the source 
was due entirely to the stellar luminosity.

For the SLS and SIS envelopes, the dust temperature is self-consistently 
calculated from the total luminosity using the dust opacity and the 
procedures described in Osorio et al.\ (1999, 2009). These authors 
calculate the dust opacity at short wavelengths ($\lambda < 200$ $\mu$m) 
assuming that the dust in the envelope is a mixture of graphite, 
silicates, and water ice, with abundances taken from D'Alessio (1996), and 
assuming a power law of the form $\kappa_\lambda \propto 
\lambda^{-\beta}$, with $1\le \beta \le 2$, for $\lambda \ge 200$ $\mu$m.

The temperature in the TSC case is also self-consistently calculated from 
the total luminosity, following the procedures described in Calvet et al. 
(1994) and Osorio et al. (2003). The latter authors obtain the dust 
opacity over the whole wavelength range assuming a mixture of graphite, 
silicates and water ice, whose parameters (grain size and abundance) are 
obtained by fitting the well sampled SED of the prototypical class I 
object L1551 IRS5.

We computed the SEDs of the models and compared them with the observed 
values of the flux density of HH 80N-IRS1 (Table~\ref{flux}). 
Additionally, we produced synthetic maps of the model emission at 3.5~mm, 
1.2~mm, and 350~$\mu$m bands. In order to fully simulate the observations, 
the synthetic maps at 1.2 mm and 350~$\mu$m were convolved with Gaussians 
with FHWM of $10.5\arcsec$ and $8.5\arcsec$, respectively. For the 
synthetic map at 3.5~mm, the effect of the missing short spacings of the 
interferometric observations was taken in account. We used the UVMODEL 
task of MIRIAD to compute the visibility tables of the 3.5 mm models with 
the same $u$-$v$ plane coverage as our PdBI observations. Then, from the 
model visibility tables, we obtained synthetic maps following the standard 
data reduction routines of the MIRIAD package.

Finally, to check the goodness of our results, we compared selected models 
with the data by means of spatial intensity profiles obtained using the 
task CGSLICE of the MIRIAD package, for both the synthetic and observed 
maps at 1.2~mm, 3.5~mm, and 350~\micron. We present two cuts of the 
observed intensities with P.A.~$\simeq 120\arcdeg$ and $\simeq 30\arcdeg$ 
(mainly along the major and minor axes of the HH 80N core seen at 1.2~mm 
and 350~\micron). An intensity profile obtained with a cut with a selected 
PA, instead of averaging the emission over an annulus, prevents to include 
contamination of the Southeastern Condensation. In practice, the synthetic maps 
of the models in the three approaches have radial symmetry since, in the 
TSC case, the angular scale of the flattening of the envelope is small 
enough that it only affects the central pixel of the map. Therefore, for the 
synthetic maps, we obtained a cut along the diameter of the modeled 
source.
 
We explore the SLS and SIS cases separately by running a grid of models 
taking the mass infall rate (\accratei), mass of the central embedded 
object (\mstar) and the external radius of the core (\rext) as free 
parameters. Because of the complex morphology of the source (see 
Fig.~\ref{cont_ammonia}) we do not adopt a fixed value of \rext; rather, 
its value is constrained in the fit. The opacity index ($\beta$) was 
derived to get a trade off to reproduce simultaneously the emission at 
1~mm and 350~\micron. This yields $\beta \sim 1.6$ for the SLS case and 
$\beta \sim 1.1$ for the SIS case. For the stellar radius ($R_{*}$) we 
chose a standard value of 5~$R_{\odot}$ (Schaller et al.\ 1992). Given the 
space of parameters (\rext, \accratei), we tested values of $M_{*}$ from 
0.5 to 1~\msun. Because a fraction of the luminosity is due to infall, 
higher values of \mstar\ would yield luminosities that exceed the upper 
limit of 110~\lsun\ derived above, assuming that \accratei~$> 10^{-5} 
M_{\odot}$~yr$^{-1}$.

The best fit model can be determined by calculating the \chisq-statistics 
obtained from the residual map resulting by subtracting the synthetic from 
the observed maps. This process was performed for all the sub/mm bands. To 
find the best fit to the images, we do not include contamination of 
the Southeastern Condensation in the \chisq\ analysis. For the 350~\micron\ 
band, the \chisq\ function was calculated over a region comprising only 
the western part of the HH 80N core, roughly a semicircle with a radius of 
$\sim 20\arcsec$. In the 1.2~mm map, HH 80N-IRS1 cannot be separated from 
the Southeastern Condensation due to the lower angular resolution of the IRAM 
30m observations with respect to the APEX observations. For this case we 
fit the intensity profile obtained along the minor axis of the emission 
(i.e. towards NE). For the 3.5~mm map, as HH 80N-IRS1 appears detached 
from the Southeastern Condensation in the map, a box of $ \sim 5 \arcsec \times 
10 \arcsec$ enclosing the entire source was used. Additionally, we 
calculated the \chisq\ function for the SED by comparing the 
fluxes of Table~\ref{flux} with the predicted SED of the model.

Because the TSC models predict the development of a flattened rotating 
structure at the innermost part of the envelope, we carried out the 
modeling assuming a TSC envelope falling onto a disk surrounding the central object.
We browsed an appropriate disk model from the online catalog of models of 
irradiated accretion disks around pre-main sequence stars (D'Alessio et 
al.\ 2005). For consistency, the orientation of the disk is chosen to 
coincide with the rotation axis of the TSC envelope and the disk radius is 
fixed to the value of the centrifugal radius $R_c$. To obtain the SED of 
the total composite model, we added the fluxes of the disk and envelope at 
each wavelength accounting for the extinction of the envelope, which is 
important at short wavelengths. Since the disk is unresolved even in the 
images with the highest angular resolution, to obtain the synthetic maps 
we added the disk to the envelope as a point source at the central pixel.

Due to computational limitations, the fitting process of the TSC envelope 
plus circumstellar disk could not be automated. Our strategy, then, was
to  perform a case-by-case exploration varying the external radius of the
core  (\rext), the radius of the expansion wave (\rew), and the reference
density ($\rho_1$) (this latter parameter is the density the envelope
would have at a radius of 1~AU for the limit $R_c = 0$, and it is related
to the mass infall rate and the central mass through equation 3 of Kenyon
et al. 1993),  until we found a model that explains satisfactorily the SED
and the  intensity profile at 350~\micron, 1.2~mm and 3.5~mm. The caveat
of this  method is that there is no assurance of finding a unique best
fit  model. However, the goal of this section is to prove that the
observed  properties of the continuum emission observed in the HH 80N core
can be  explained in terms of a protostar plus an infalling envelope that
is  embedded inside the HH 80N core. Constraining a unique model with a 
flattened envelope and a disk requires additional mid-IR observations
with  high angular resolution and it is beyond the scope of this paper.

\subsection{Results for the SLS model}

\begin{figure}[fh]
\epsscale{0.95}
\plotone{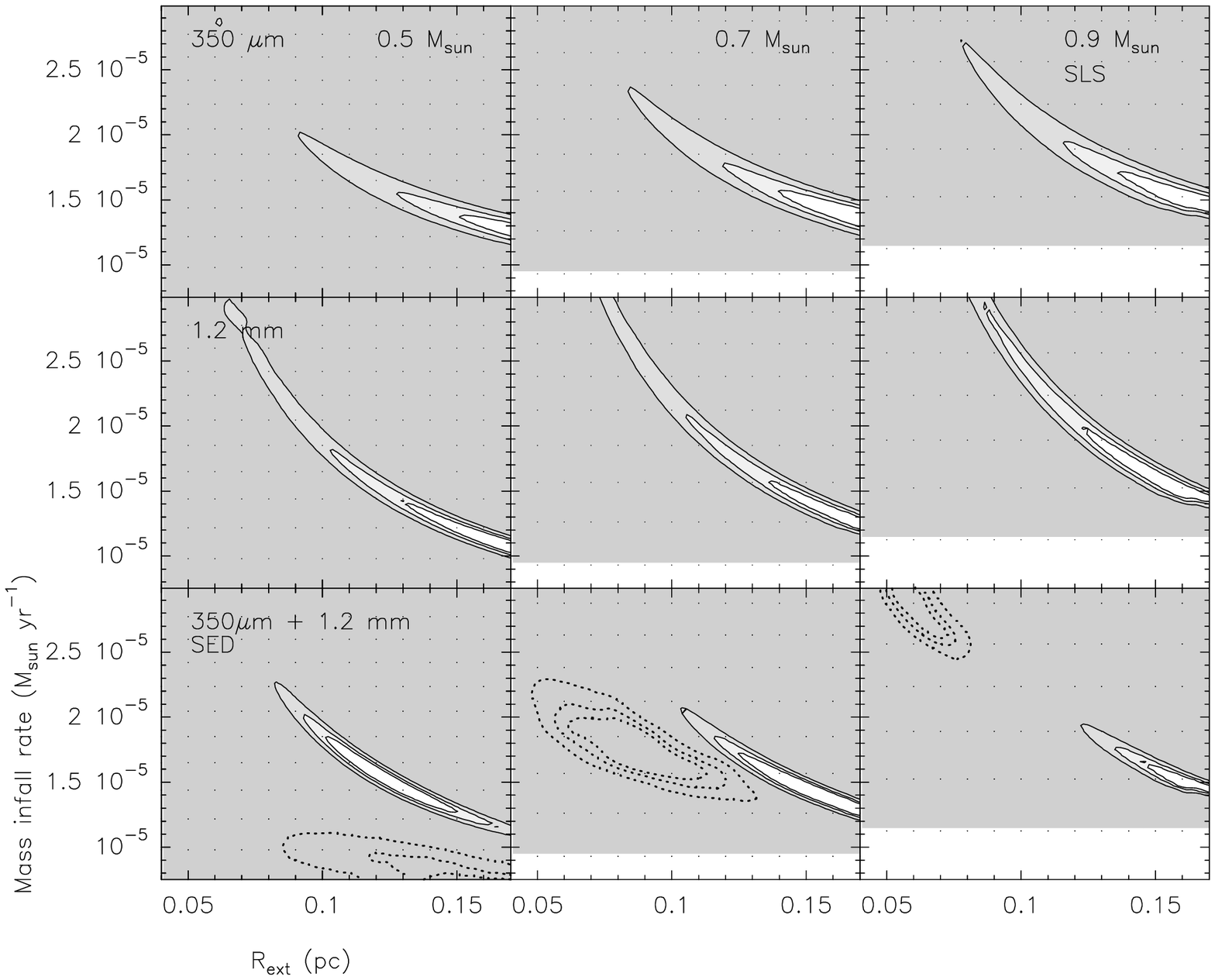}
\vspace{0.0 cm}
\caption{
Contour plots (solid line and gray scale) of the \chisq\ function derived for the fitting of the models of the collapse of the Singular
Logatropic Sphere to the observed images. Rows correspond to the 350~\micron\ band (top), the 1.2~mm band (middle), and the sum of the \chisq\
function for the 350~\micron\ and 1.2~mm bands (bottom). The dotted contours shown in the bottom panels correspond to the \chisq\ function
derived for the SED. Columns correspond to \mstar\ = 0.5 \msun\ (left), 0.7 \msun\ (middle), and 0.9 \msun\ (right). In all the cases the
contour levels correspond to the confidence levels of 99\%, 90\% and 68\% (1$\sigma$). These contours are relative to the minimum \chisq\ value of
each panel, which may vary significantly for different masses.  For instance, in the bottom row, where we compare the results of the SED and
single-dish maps, the minimum \chisq\ values of the SED are 53.4, 33.2, and 19.2 for 0.5, 0.7, and 0.9~\msun, respectively; the sum of the
minimum \chisq\ values of the 350~\micron\ and 1.2~mm maps are 102.2, 119.0, and 122.2 for 0.5, 0.7, and 0.9~\msun, respectively.
\label{chi2_log}}
\end{figure}

We tested the logatropic density distribution by using 1080 different  models within
the following set of ranges for the external radius, mass  infall rate and mass of
the central embedded object: 0.04~pc $\leq$ \rext\  $\leq$ 0.18~pc, $7 \times
10^{-6} M_{\odot}$~yr$^{-1} \leq$~\accratei~$\leq  5 \times 10^{-5}
M_{\odot}$~yr$^{-1}$ and 0.5~\msun\ $\leq$ \mstar\ $\leq$  1.0~~\msun. In this space
of parameters we expect to find meaningful  physical solutions. To establish the
goodness of the fit we calculated the  \chisq\ as discussed in the previous section.
Figure~\ref{chi2_log} shows  the best set of solutions for the \chisq\ estimated
separately for the SED  and for the 1.2~mm and 350~\micron\ intensity distributions.
We also  calculated the \chisq\ function for the 3.5~mm band but the results are 
not included in the figure because no set of parameters can provide reasonable
\chisq\ values at this band. We consider as good solutions those where the
\chisq\ values of the  intensity distributions and of the SED are all within the 90\%
level of  confidence.

\begin{figure}[fh]
\epsscale{1.0}
\plotone{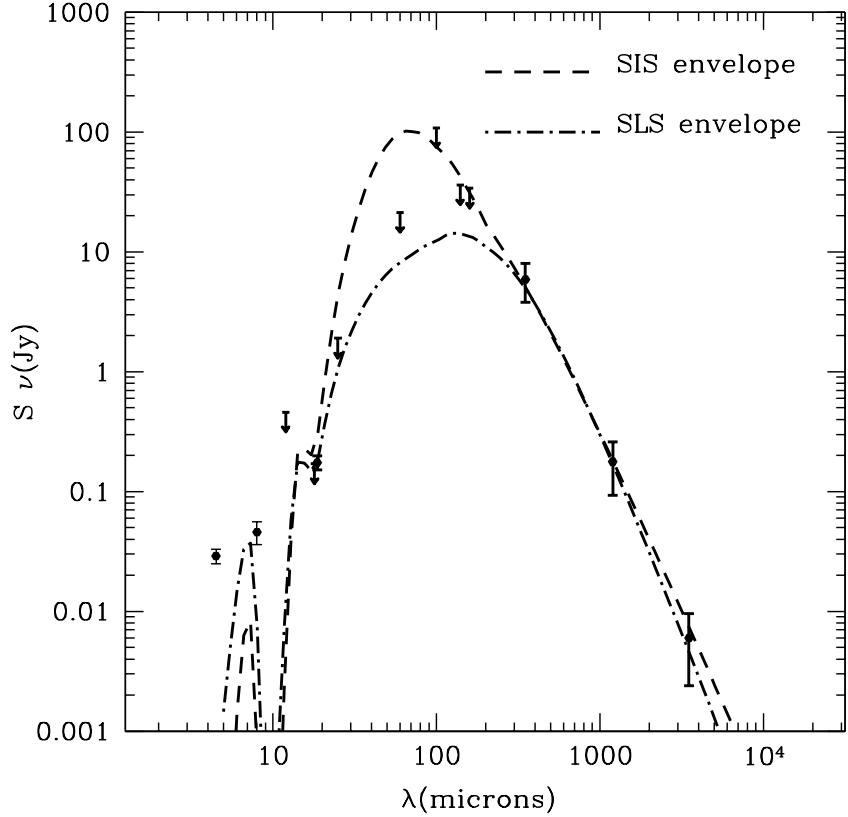}
\vspace{0.0 cm}
\caption{
Observed flux densities (dots; see Table~\ref{flux}) and predicted SEDs for HH 80N-IRS1, assuming models of the collapse of the Singular
Logatropic Sphere (SLS; dash-dotted line) and of the Singular Isothermal Sphere (SIS; dashed line). Bars represent the uncertainties and
arrows represent upper limits. The SLS model corresponds to the model that minimizes the \chisq\ of the SED (\rext\ = 0.1~pc, \mstar =
0.7~\msun, \accratei = 1.57~$\times 10^{-5}~M_{\odot}$~yr$^{-1}$, $L_\mathrm{bol} = 69$~\lsun, $M_\mathrm{env} = 25.5$~\msun, and $\beta$~=~1.6). The SIS
model corresponds to the model that minimizes the \chisq\ function for the maps at 1.2~mm and 350~\micron\ (\rext~=~0.05~pc,
\mstar~=~0.8~\msun, \accratei~=~1.05~$\times 10^{-4} M_{\odot}$~yr$^{-1}$, $L_\mathrm{bol} = 527$~\lsun, $M_\mathrm{env} = 4.6$~\msun, and $\beta$~=~1.1).
\label{sed_esf}}
\end{figure}

\begin{figure}[fh]
\epsscale{1.0}
\plotone{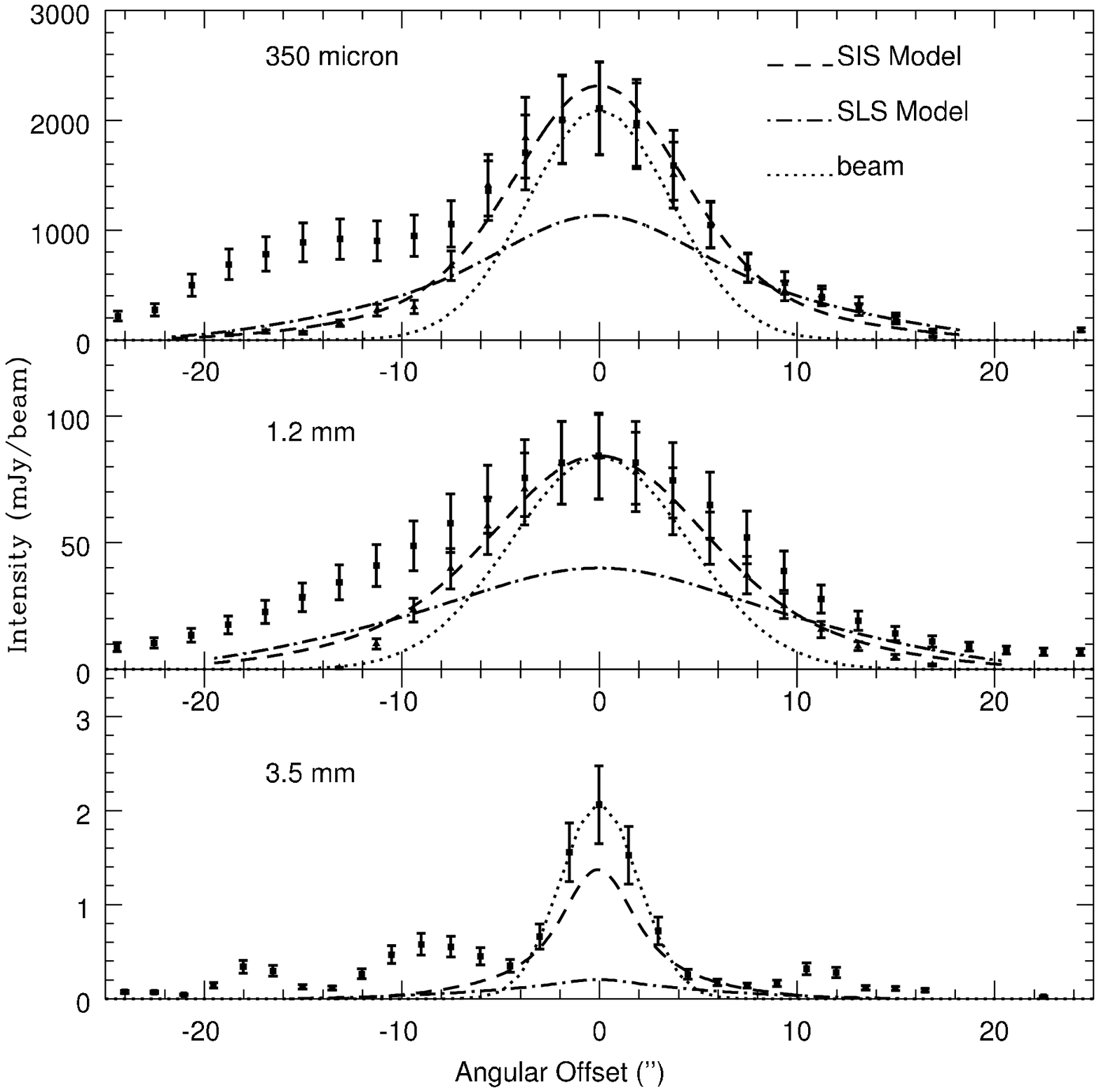}
\vspace{0.0 cm}
\caption{
Observed and modeled intensity profiles of HH 80N-IRS1 at 350~\micron\ (top), 1.2 mm (middle) and 3~mm (bottom). The squares and triangles represent observed cuts along the major (P. A.~$\simeq 120 \arcdeg$) and minor (P. A.~$\simeq 30 \arcdeg$) axes of the HH 80N core, respectively. Bars represent the uncertainties.  At 3.5~mm, as the source appears unresolved, we present only a cut along the major axis of the core (P. A.~$\simeq 120 \arcdeg$). The dash-dotted lines correspond to cuts along the diameter of the synthetic maps of the Singular Logatropic Sphere model that minimizes the \chisq\ for the SED (see  Fig.~\ref{sed_esf}). The dashed lines correspond to cuts along the diameter of the synthetic maps of the Singular Isothermal Sphere model that minimizes the \chisq\ for the maps (see  Fig.~\ref{sed_esf}). The dotted line represents the beam.
\label{perfils_esf}}
\end{figure}

Figure~\ref{chi2_log} shows that there are no solutions that fit together 
the SED and the intensity distribution of the 1.2~mm and 350~$\mu$m maps 
(i.e., there is no overlap between the \chisq\ contours of the maps and 
those of the SED).  As an example, Figures~\ref{sed_esf} and 
\ref{perfils_esf} show, respectively, the predicted SED and intensity 
profiles (dash-dotted lines) of the SLS model that gives the minimum 
\chisq\ for the SED, compared with the observed data. Except for the 
mid-IR wavelengths, the observed SED is reasonably well reproduced by the 
model. However, as Figure~\ref{perfils_esf} shows, this model produces 
intensity profiles too flat and cannot reproduce the observed intensity 
profiles in any of the bands. Given these discrepancies, we conclude that 
in the logatropic case the mass is not distributed adequately in the 
envelope to match the observations.

\subsection{Results for the SIS model} 

\begin{figure}[fh]
\epsscale{0.95}
\plotone{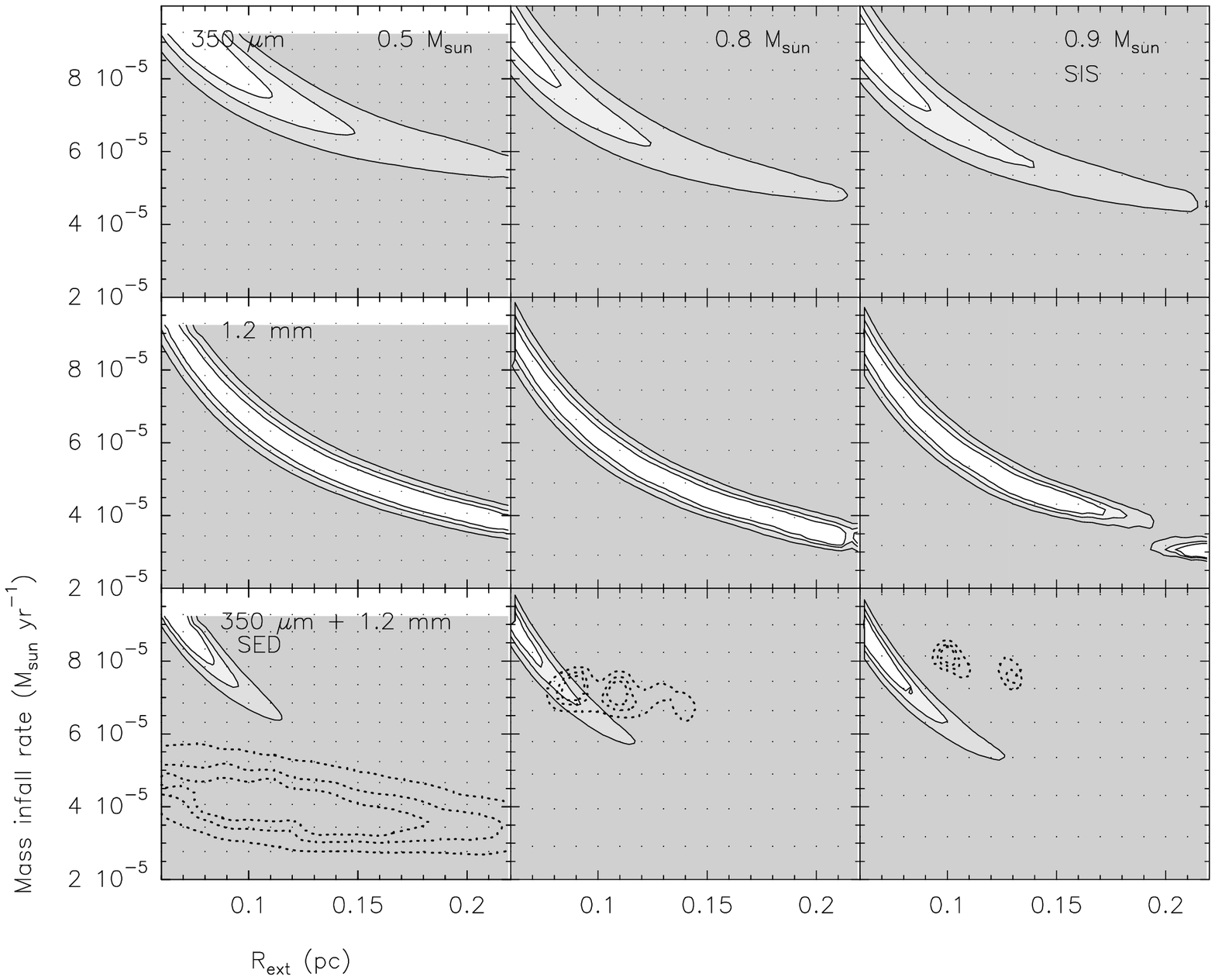}
\vspace{0.0 cm}
\caption{
Contour plots (solid line and gray scale) of the \chisq\ function derived for the fitting of the models of the collapse of the Singular
Isothermal Sphere to the observed images of HH 80N-IRS1. Rows correspond to the 350~\micron\ band (top), the 1.2~mm band (middle), and the sum
of the \chisq\ function for the 350~\micron\ and 1.2~mm bands (bottom). The dotted contours shown in the bottom panels correspond to the
\chisq\ function derived for the SED. Columns correspond to \mstar\ = 0.5 \msun\ (left), 0.8 \msun\ (middle), and 0.9 \msun\ (right).  In all
the cases the contour levels correspond to the confidence levels of 99\%, 90\% and 68\% (1$\sigma$). These contours are relative to the
minimum \chisq\ value of each panel, which may vary significantly for different masses.  For instance, in the bottom row, where we compare the
results of the SED and single-dish maps, the minimum \chisq\ values of the SED are 47.1, 278.6, and 600.3 for 0.5, 0.7, and 0.9~\msun,
respectively; the sum of the minimum \chisq\ values of the 350~\micron\ and 1.2~mm maps are 40.6, 36.1, and 36.7 for 0.5, 0.7, and 0.9~\msun,
respectively. \label{chi2_iso}
}
\end{figure}

We tested the isothermal density distribution by using 1800 different 
models within the following set of ranges for external radius, mass infall 
rate and mass of the central embedded object: 0.03~pc $\leq$ \rext $\leq$ 
0.22~pc, $1.2 \times 10^{-5} M_{\odot}$~yr$^{-1} \leq$~\accratei~$\leq 1.6 
\times 10^{-4} M_{\odot}$~yr$^{-1}$ and 0.5~\msun\ $\leq$ \mstar\ $\leq$ 
1.0~~\msun. Figure~\ref{chi2_iso} shows the best models from the \chisq\ 
analysis for the SIS case. For the same reasons as in the logatropic case, 
we have not included the 3.5~mm results in the figure.

The mass infall rates considered for the SIS collapse are higher  than those considered
in the logatropic case. This is because, for a given value of the mass infall rate, the
SIS models yield less massive envelopes than the SLS models; so, in order to fit 
properly the mm flux densities higher values of the mass infall rate are required in
the SIS models (see Osorio et al. 1999). For this reason,  \rext\ becomes almost
irrelevant and \accratei\ takes the dominant role in  the fitting.
Figure~\ref{chi2_iso} (bottom panels) shows that for the  \mstar~$= 0.8$~\msun\ case,
there are solutions that apparently satisfy  both the SED and (sub)mm intensity
distribution constraints (i.e., for 0.8~\msun, the \chisq\  contours of the SED overlap
those of the maps for  \rext~$\simeq$ 0.08-0.1~pc and \accratei~=~(6.5-8.0)$ \times
10^{-5}~M_{\odot}$~yr$^{-1}$). Nevertheless, despite these solutions  yielding
reasonable intensity distributions in the 1.2~mm and 350~\micron\  bands, they
overestimate the total luminosity, specially at far-IR  wavelengths. In
Figure~\ref{sed_esf} (dashed line) we show an example that illustrates this behavior
(the predicted flux is almost one order of magnitude above the  observed flux at
60~\micron). As in the SLS case, the modeled intensity profile at 3.5~mm (dashed line
in Fig.~\ref{perfils_esf}) is significantly weaker than the observed profile. These
results are general and we can find solutions that can fit  the single--dish intensity
distribution of the envelope but they predict  an excess of luminosity and fail to
reproduce the compact emission seen in  the 3.5 mm PdBI map.  In conclusion, as in the
SLS case, there is no SIS model  that can fit the SED and the maps simultaneously.

\subsection{Results for the TSC model}

One of the caveats of the SLS and SIS models is that they are unable to fit the intensity
of the compact source seen in the PdBI 3.5~mm map. This compact source could have a
significant contribution from a circumstellar disk. In this section we model the source
assuming a TSC envelope falling onto a disk surrounding the central object obtained from
the catalog of D'Alessio et al.\ (2005). Apart from using a more realistic model, such a
configuration is ideal for two reasons. First, if the circumstellar disk is populated by
millimeter-size grains, its emission can be significant at mm wavelengths and, because it
is compact, it will be less affected by the interferometric filtering that affects the
envelope emission. Second, because the SED at short wavelengths is very sensitive to the
geometry of the source, the TSC envelope and a disk with the proper inclination can
provide the extinction required to fit the mid-IR part of the SED, depending on the
inclination of the axis of the system (envelope plus disk). This would give luminosities
similar to or below 110~\lsun. In addition, the infall luminosity is reduced because
the material lands on a disk instead of falling directly onto the protostar.

\begin{figure}[fh]
\epsscale{1.0}
\plotone{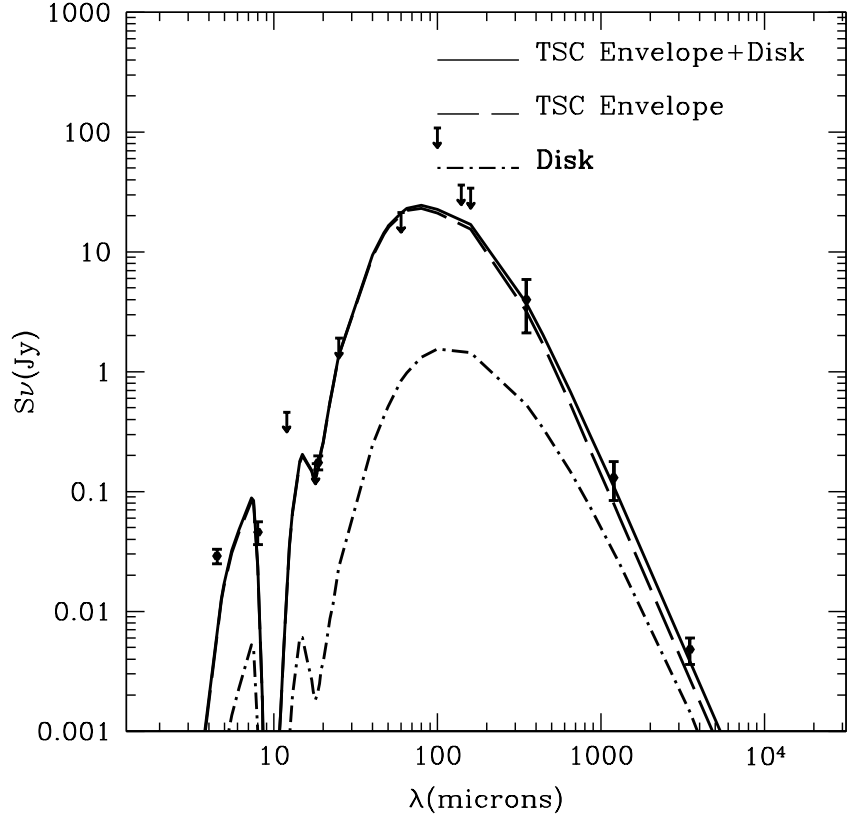}
\vspace{0.0 cm}
\caption{
Observed flux densities (dots; see Table~\ref{flux}) and predicted SED for HH 80N-IRS1, assuming the Terebey, Shu, and Cassen (TSC) envelope
model plus an accretion disk. Bars represent the uncertainties and arrows represent upper limits. The dashed line represents the SED of our
favored TSC envelope model ($R_{\rm ext} = 1.8 \times 10^{4}$~AU,  \mstar = 3~\msun, \accratei~$\simeq 1.6 \times 10^{-4}
M_{\odot}$~yr$^{-1}$, $L_\mathrm{bol} = 105$~\lsun, and  $M_\mathrm{env} = 20$~\msun). The point dashed line represents the SED of the selected disk model
(\accrateacc~$=10^{-7} M_{\odot}$~yr$^{-1}$, $R_c =$~300~AU, and $i = 30\arcdeg$). The solid line represents the resulting SED of out
favored TSC envelope plus disk model. 
\label{sed_tsc}}
\end{figure}

\begin{figure}[fh]
\epsscale{1.0}
\plotone{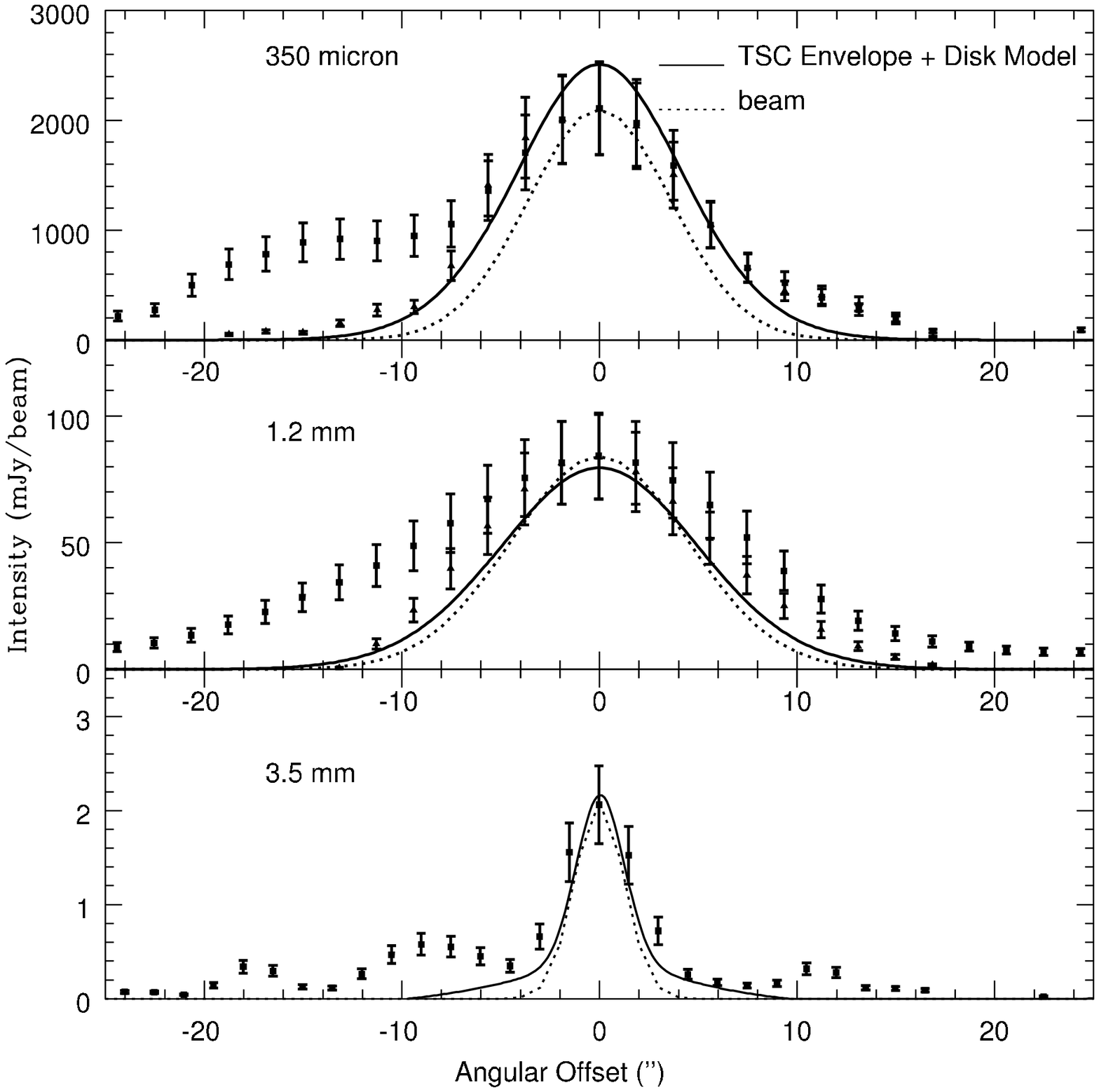}
\vspace{0.0 cm}
\caption{
Observed and modeled intensity profiles of HH 80N-IRS1 at 350~\micron\ (top), 1.2 mm (middle) and  3~mm (bottom). The squares and triangles represent observed cuts along the major (P. A.~$\sim 120\arcdeg$) and minor (P. A.~$\sim 30\arcdeg$) axes of the HH 80N core, respectively. Bars represent the uncertainties. 
At 3.5~mm, as the source appears unresolved, we present only a cut along the major axis of the core (P. A.~$\sim 120\arcdeg$). The solid lines correspond to cuts along the diameter of the synthetic maps of our favored Terebey, Shu, and Cassen envelope plus disk model (see Fig.~\ref{sed_tsc}).
The dotted line represents the beam. 
\label{perfils_tsc}}
\end{figure}

We tested several TSC envelopes exploring values of the central luminosity  
$50 < L_* <  250$~\lsun, outer radius of the envelope 
 %$1.5 \times 10^{4}$ and $3.8 \times 10^{4}$~AU 
$0.07 < R_{\rm ext} < 0.18$ pc, the radius of the expansion wave $0.03 < R_{\rm ew} <
0.09$ pc, and reference density $4.1\times
10^{-13} < \rho_1 < 1\times  10^{-12}$~g~cm$^{-3}$ (these values of $\rho_1$ are
equivalent to densities at 1000 AU between $7.8 \times 10^{5}$ and $7.4 \times
10^{6}$~cm$^{-3}$, and correspond to values of the mass infall rate between $3.2\times
10^{-5}$ and $3.3\times 10^{-4}~M_\odot$~yr$^{-1}$ for a 3~$M_\odot$ star). For the
circumstellar disk we explored accretion rates from the disk to protostar between
$10^{-9}$ to $10^{-6}  M_{\odot}$~yr$^{-1}$, and we assume that the disk is irradiated
with a similar luminosity and has a similar inclination than the envelope. In
Table~\ref{TSC} we give the parameters of our favored model. Figures~\ref{sed_tsc} and
\ref{perfils_tsc} show the observed SED and intensity profiles, respectively, predicted
by  our favored TSC model. These figures show that the SED is well reproduced  by the
model in almost all the data points and that the modeled intensity  profiles fit
reasonably well the observations within the calibration  uncertainties.

 At 3.5~mm, the inclusion of a disk provides the flux needed  to explain the observed
intensity peak. We note that, possibly, a fine tuning of the mass accretion rate could
provide a more accurate fit for the intensity profiles. However, the range of accretion
rates of the online disk model  grid is sampled in steps that vary one order of
magnitude and the next  available model has an accretion rate too small and provides
too faint mm  emission. Nevertheless, in this analysis we aimed to prove that our 
continuum observations can be explained in the frame of standard star  forming models
and our favored model presented above fulfills this  requirement. Table~\ref{Tchi2}
shows a summary of the \chisq\ analysis (SED,  single-dish and interferometric maps)
for this TSC (+disk) model (Figs.~\ref{sed_tsc},  \ref{perfils_tsc}), as well as for
the SLS and SIS models shown in Figures~\ref{sed_esf},  \ref{perfils_esf}. The reduced
\chisq\ values show that in the TSC case we  obtain a better fit.

 \section{Discussion}
 
The derived physical parameters of the selected TSC model listed in Table~\ref{TSC}
suggest that HH~80N-IRS1 is a very young Class 0 protostar. First, the predicted  volume
density for HH~80N-IRS1 at 1000~AU, $n({\rm H_2}) \simeq 4 \times 10^{6}$~cm$^{-3}$, is
typical of Class 0 sources (J\o rgensen et al.\ 2002), and greater than Class I sources
(J\o rgensen et al.\ 2002) and prestellar cores (Kirk, Ward-Thomson \& Andr\'e\ 2005;
Tafalla et al.\ 2002); second, the estimated upper limit of the mass infall rate, $\sim 2
\times 10^{-4} M_{\odot}$~yr$^{-1}$ (see Table~\ref{TSC}), is compatible with the high
values of the mass infall rate typical of  young Class 0 protostars (Maret et al.\ 2002),
and yields to an age of $\sim 2\times 10^{4}$~yr for HH~80N-IRS1; finally, HH 80N-IRS1
fulfills the criteria proposed by Andr\'e et al.\ (1993) for Class 0 objects, $L_{\rm
submm}/L_{\rm bol}\ga 5\times 10^{-3}$, which in our case is about 0.1.  On the other
hand, the derived luminosity of 105~\lsun\ for HH~80N-IRS1 is found in the threshold
between low mass and intermediate mass protostars. 
Given the large reservoir of mass of
the HH 80N core (see 1.2~mm and 350~\micron\ map of Fig.~\ref{cont_ammonia}) and the
youth of the HH~80N-IRS1, we cannot discard further accumulation of material towards the
central object.  

From the integrated emission of the residual map at 1.2~mm (resulting from the
subtraction of the emission of the synthetic map of the TSC model from the map
observed with MAMBO) we can obtain a crude estimate of the mass of the HH 80N
core outside the HH 80N-IRS1 envelope. Assuming optically thin dust emission
with $\beta = 2$ and a temperature of $\sim$14~K (the boundary temperature of
the HH 80N-IRS1 envelope) we derive a mass of $\sim$10~\msun\ for this material.
As we noted above, the HH 80N core, with a size of $0.16 \times 0.12$~pc and an
estimated total mass of $\sim$30~\msun\ (20~\msun\ of HH 80N-IRS1 + 10~\msun\ of
the rest of the HH 80N core), contains more material than the infalling envelope
associated with IRS1. According to the results of our TSC modeling (see
Table~\ref{TSC}) the mass of the envelope is 20 $M_\odot$, and the infall occurs
within a radius of \rew\ = $1.5 \times 10^4$~AU with the envelope being static
outside this radius. Furthermore, the molecular emission of tracers such CS, SO
and HCO$^+$ extends over a region considerably larger than the HH~80N core, as
traced by the dust continuum and ammonia line emissions.  Indeed, the emission
of these molecular tracers has been interpreted as arising from a contracting
ring around HH~80N-IRS1 with an inner radius (the radius of the region where
these molecular species appear to be depleted) of $2.5 \times 10^4$ AU and an
outer radius of $6 \times 10^4$~AU (Girart et al.\ 2001; Masqu\'e et al.\
2009).  The estimated average volume density of the molecular ring is in the
$5\times10^4$-$1.3\times10^5$~cm$^{-3}$ range (Masqu\'e et al.\ 2009), which
seems too high, given the estimated density in the static part of the
HH~80N-IRS1 envelope ($\sim 6 \times 10^4$~cm$^{-3}$, according to our
modeling). Therefore, the kinematics and physical conditions in the molecular
ring-like structure proposed by Girart et al.\ (2001) and Masqu\'e et al.\
(2009) appear puzzling. The role of the HH 80/81/80N outflow in the properties
of this molecular ring-like structure and the relationship with the onset of the
star-forming process in the HH 80N core, in particular with the HH 80N-IRS1,
protostar is an interesting issue that deserves further observational and
theoretical investigation.

\section{Conclusions}

We have carried out dust continuum and ammonia line observations of the dense core ahead of HH 80N, complemented with archive
data, covering a wide range of wavelengths. We analyzed the continuum data by means of self-consistent
models using several approaches for the envelope structure and we discuss the inclusion of a protostellar
disk. Additionally, we compare ammonia observations of the (1,1) transition with continuum emission (and
absorption) maps. Our main conclusions are summarized as follows:

\begin{enumerate}

\item The \ammonia\ (1,1) emission shows a striking correlation with the dust continuum
emission and with the absorption silhouette  seen in the 8~\micron\  Spitzer image.  This
indicates that the ammonia traces fairly well the distribution of gas and dust in the HH 80N
core. Pending a proper analysis of the \ammonia\ abundances, this preliminary assessment
points that there is no need to invoke photochemical effects caused by the nearby HH 80N
object to explain the distribution of ammonia in the HH 80N core. However, a detailed
inspection of the ammonia map shows that an important part of the \ammonia\ emission arises
from the Southeastern part of the core, close to HH 80N, which could be due to a slight
abundance enhancement.

\item The continuum emission presents a peak at the same position
($\alpha(J2000)=18^\mathrm{h}19^\mathrm{m}17\fs81$, $\delta(J2000)=-20\arcdeg40'47\farcs7$) in all the
bands (4.5~\micron, 8~\micron, 350~\micron, 1.2~mm and 3.5~mm). This emission peak is located at the
center of the CO bipolar outflow found by Girart et al.\ (2001), suggesting the presence at this position of an embedded young stellar object (HH 80N-IRS1) that powers the outflow.

\item We find that the SED and the intensity distribution of the mm and submm emission of HH 80N-IRS1 can be reproduced by
a slowly rotating infalling envelope described by the Terebey, Shu, and Cassen (TSC) solution, plus a circumstellar
accretion disk. The mass of the envelope is 20~\msun, the central luminosity is 105~\lsun, and the radius of the infalling
region is 1.5$ \times 10^4$~AU. The disk has a mass of 0.6~\msun\ and a radius of 300 AU. Such a configuration, together
with the derived high values of the mass infall rate ($1.7 \times 10^{-4}\,(M_*/3\,M_{\odot})^{1/2}~M_{\odot}$~yr$^{-1}$),
and  young age ($\sim 2\times 10^{4}$~yr), suggest the HH 80N-IRS1 may be a young Class 0 source. 

\item The APEX map at 350~\micron\ and, especially, the PdBI map at 3.5~mm, where the extended emission is resolved out,
show signs of possible fragmentation suggesting that other sources, in addition to HH 80N-IRS1, could be embedded inside
the HH 80N core. On the other hand, previous studies reveal that the molecular emission of some tracers is considerably
more extended than the dust and \ammonia\ emission presented in this work. This suggests that the HH 80N core is surrounded
by a larger molecular structure whose properties could be influenced by the proximity of the HH 80/81/80N outflow.

\end{enumerate}

\acknowledgments

G.A., R.E., J.M.G., J.M.M., and M.O. acknowledge support from MICINN (Spain) grant
AYA2008-06189-C03 (co-funded with FEDER funds). G.A. and M.O. acknowledge partial support
from Consejer\'\i a de Innovaci\'on, Ciencia y Empresa de la Junta de Andaluc\'{\i}a
(Spain). G.G. acknowledges support from CONICYT projects FONDAP No. 15010003 and BASAL
PFB-06. We thank Susana Lizano for providing us the routines for the calculation 
of the logatropic density distribution.

\clearpage

\clearpage

\clearpage

\clearpage

\clearpage

\clearpage

\clearpage

\clearpage

\clearpage

\begin{table}
\caption{Positions of the compact Spitzer sources\tablenotemark{a}\label{spitzer}}
\begin{tabular}{lcc}
\hline\hline
                             &   \multicolumn{2}{c}{Peak Position}                                                                                 \\                                        
Source                       &                  RA(J2000)                                               &       DEC (J2000)                          \\ 
\hline
HH 80N-IRS1         &   $18^\mathrm{h}19^\mathrm{m}17\fs81$          &     $-20\degr40'47\farcs7$           \\
HH 80N-IRS2         &     $18^\mathrm{h}19^\mathrm{m}19\fs01$        &      $-20\degr40'56\farcs4$      \\
HH 80N-IRS3        &     $18^\mathrm{h}19^\mathrm{m}17\fs42$        &       $-20\degr40'34\farcs3$           \\
\hline 
\end{tabular}
\tablenotetext{a}{HH 80N-IRS1 position was derived from the 8~\micron\ image.
HH 80N-IRS2 and HH 80N-IRS3 positions were derived from the 4.5~\micron\
image. 
}
\end{table}

\begin{table}
\scriptsize
\caption{Source parameters derived from the PdBI 3.5~mm map\tablenotemark{a} \label{sources}}
\begin{tabular}{lcccccc}
\hline\hline
                             &   \multicolumn{2}{c}{Peak Position}                                                                            &         $I_\nu$(peak)\tablenotemark{b}     &       $S_{\nu}$\tablenotemark{c}     &       Deconvolved Size         & P. A.         \\                                        
Source                &                  RA(J2000)                                               &       DEC (J2000)                      &      (mJy~beam$^{-1}$)      &       (mJy)                &             ($\arcsec$)                 &   ($\arcdeg$)          \\ 
  \hline
HH 80N-IRS1          &     $18^\mathrm{h}19^\mathrm{m}17\fs81$         &     $-20\degr40'47\farcs7$    &            $2.03\pm0.11$        &     $3.75\pm0.19$        &  $5.0 \times 3.0$  &    -17.4       \\
Southeastern Condensation     &     $18^\mathrm{h}19^\mathrm{m}18\fs31$       &       $-20\degr40'52\farcs2$     &           $0.58\pm0.06$        &     $2.48\pm0.09$        &  $11.1 \times 5.8$  &    -17.0    \\
\hline 
\end{tabular}
\tablenotetext{a}{Derived from a Gaussian fit with the task IMFIT of MIRIAD.}
\tablenotetext{b}{Peak Intensity.}
\tablenotetext{c}{Integrated flux density.}
\end{table}

\begin{table}
\scriptsize
\caption{Summary of the continuum data of HH 80N-IRS1\label{flux}}
\begin{tabular}{lccccccc}
\hline\hline
      &                         &   Angular      &        Aperture         &  Flux                           &   &      \\                                        
Wavelength      &               &   Resolution\tablenotemark{a}   &       Size\tablenotemark{b}         & Density\tablenotemark{c}                               &    Observing    &            \\ 
   (\mi)        &  Instrument   &  ($\arcsec$)                &   ($\arcsec$)                          &  (Jy)                       &         Epoch       & Notes \\
\hline
3500       &    IRAM~PdBI           &    2.9 $\times$ 7.0    &  $\sim 5 \times 10$ &  0.004--0.006   &           April 2010 & This paper \\
1200       &    IRAM~30m (MAMBO II)        &     10.5                  &                    &  0.084--0.177   &                       March 2008 & This paper \\
350        &    APEX (SABOCA)           &  8.5\tablenotemark{d}                    &     $\sim 30 \times 50$     & 2.1--5.9 &          October 2009  & This paper   \\
160       &  Akari  (FIS) &   $\sim$~60    &                             --        &     $\le$~33.88                                                    & 2006-2007  &Archive data \\
140      &  Akari  (FIS)  &  $\sim$~55   &                             --         &    $\le$~36.11                                                      & 2006-2007  & Archive data \\
100        &    IRAS     & 180 $\times$ 300                &            --                               &  $\le$~108  &  1983    &  Archive data   \\
60         &    IRAS     &  90 $\times$  282                 &            --     &                              $\le$~22.08                  &     1983      &    Archive data      \\
25         &    IRAS     &  45 $\times$  276                &            --       &                          $\le$~1.24               &              1983     &   Archive data    \\
%19.5         &    VLT (VISIR)     &  $\sim 0.5$                      &  1.5                                        & $\le$~8.32                        &    May 2009        &   This paper        \\
18.7         &    VLT (VISIR)     &  $\sim 0.5$                      &  1.5                                        &  0.175 (0.01)                      &    June 2009         &   This paper       \\
18.0 &  Akari (IRC)  &   $\sim$~7          & -- &      0.170 (0.009)\tablenotemark{e}                                     & 2006-2007   &  Archive data \\
%17.7         &    VLT (VISIR)     &  $\sim 0.5$                      &  1.5                                        & $\le$~17.34                       &    May 2009         &    This paper      \\
12         &    IRAS     &   45 $\times$  270                &             --                &  $\le$~0.46                                       & 1983  & Archive data \\
8          &    Spitzer (IRAC)   &     1.7         &  12                                         &  0.041~(0.010) & September 2005  & Archive data\\
4.5        &    Spitzer (IRAC)  &     1.9         &  11                                         &  0.026~(0.003)  & September 2005 & Archive data \\
\end{tabular}
\tablenotetext{a}{For the mm and submm data the reported angular resolution corresponds to the FHWM of the beam size. For far-IR and mid-IR data it corresponds to the Point Spread Function. For Spitzer it corresponds to the pixel size, which is larger than the angular resolution.}
\tablenotetext{b}{The box for the 350~\micron\ and 1.2~mm data has a P.A. of 120\arcdeg\ and it is chosen to include only HH 80N-IRS1 (see \S~3.1). The values given for IR data are the diameter of a circular aperture.}
\tablenotetext{c}{For the 350~\micron, 1.2~mm and 3.5~mm measurements we give a range in order to account for contamination effects from the Southeastern Condensation (see \S~3.3). For the mid infrared points uncertainties are included into parenthesis.
We adopted the IRAS and Akari flux values as upper limits because of possible contamination by background sources.}
\tablenotetext{d}{After smoothing with a Gaussian of FWHM = $4\farcs0$.}
%\tablenotetext{e}{Virtual pixel size $10 .4 \times 9 .36\arcsec$ (check!)}
\tablenotetext{e}{Adopting a 5\% of calibration uncertainty as indicated in the IRC Data User Manual.}
\end{table}

\begin{table}
\caption{Results of the TSC modeling\label{TSC}}
\begin{tabular}{lcc}
\hline\hline
Envelope parameter    &  Symbol & Value                                            \\                                        
\hline
Mass     &  $M$    &    20~$M_{\odot}$              \\ 
Central luminosity  &$L_*$   &   105~$L_{\odot}$  \\
Radius of the expansion wave\tablenotemark{a} & $R_{\rm ew} $ &  $1.5 \times 10^4$~AU  \\
Outer radius   & $R_{\rm ext}$  & $1.8 \times 10^4$~AU   \\
Inclination angle & $i$   &  30$\arcdeg$   \\
Centrifugal radius  &$R_c$ &  300~AU      \\
Reference density &$\rho_1$ & $5 \times 10^{-13}$~gr~cm$^{-3}$ \\
Density at $r = 1000$ AU & $n$~(1000 AU) & $4.1 \times 10^{6}$~cm$^{-3}$ \\
Mass infall rate \tablenotemark{b}  &$\dot M_{\rm i}$  & $1.6 \times 10^{-4} M_{\odot}$~yr$^{-1}$  \\
\hline
Disk parameter\tablenotemark{c}    & Symbol   & Value         \\
\hline
Mass &  $M_{\rm disk}$  &  0.6~$M_{\odot}$        \\
Radius\tablenotemark{d} & $R_{\rm disk}$ & 300 AU \\
Inclination angle\tablenotemark{e} & $i$  &  30$\arcdeg$   \\
Mass accretion rate & $\dot M_{\rm acc}$     & $10^{-7} M_{\odot}$~yr$^{-1}$  \\
Viscosity parametrization    & $\alpha$ &  0.01    \\
Slope of grain size distribution & p    &   3.5      \\
Minimum grain size     &  $a_{\rm min}$  &       0.005~\micron      \\
Maximum grain size     &   $a_{\rm max}$ &      1~mm        \\
\end{tabular}
\tablenotetext{a}{Radius of the infalling region. Outside this radius the envelope remains static.}
\tablenotetext{b}{Obtained adopting a mass of $M_*=3~M_{\odot}$ for the embedded
protostar (i.e. the infall rate value is an upper limit).}
\tablenotetext{c}{Obtained assuming that the disk is irradiated by a luminosity equal to the central luminosity $L_* = 105~L_\odot$.}
\tablenotetext{d}{Assumed to coincide with the centrifugal radius derived for the envelope.}
\tablenotetext{e}{Assumed to coincide with the inclination angle derived for the envelope.}
\end{table}

\begin{deluxetable}{lccc}
\tablewidth{0pc}
\tablecaption{Reduced $\chi^2$ Results}
\tablehead{
\colhead{} & 
\colhead{} & 
\colhead{Single-dish} &
\colhead{Interferometric}
\\
\colhead{} & 
\colhead{SED} & 
\colhead{maps\tablenotemark{a}} &
\colhead{map\tablenotemark{b}}
}
\startdata
SLS	model		& 34.1 	& 31.5	& 68.3\\
SIS	model		& 441.4 	& 3.3		& 13.3\\
TSC+disk model	& 3.5 	& 8.1 	& 5.5 \\
\enddata 
\label{Tchi2}
\tablenotetext{a}{1.2~mm map obtained with MAMBO at IRAM 30 m telescope, and 350~$\mu$m map obtained with LABOCA at APEX 12 m telescope.}
\tablenotetext{b}{3.5~mm map obtained with PdBI.}
\end{deluxetable}

\end{document}